# Pre-treatment radiotherapy dose verification using Monte Carlo doselet modulation in a spherical phantom

**Reid W Townson**[1,2] **and Sergei Zavgorodni**[2,1]

[1] Department of Physics and Astronomy, University of Victoria, Victoria, British Columbia, Canada

[2] Department of Medical Physics, BC Cancer Agency, Vancouver Island Centre, Victoria, British Columbia, Canada

Email: rtownson@uvic.ca

**Abstract**

Due to the increasing complexity of radiotherapy delivery, accurate dose verification has become an essential part of the clinical treatment process. The purpose of this work was to develop an electronic portal image (EPI) based pre-treatment verification technique capable of quickly reconstructing 3D dose distributions from both coplanar and non-coplanar treatments. The dose reconstruction is performed in a spherical water phantom by modulating, based on EPID measurements, pre-calculated Monte Carlo (MC) doselets defined on a spherical coordinate system. This is called the spherical doselet modulation (SDM) method. This technique essentially eliminates the statistical uncertainty of the MC dose calculations by exploiting both azimuthal symmetry in a patient-independent phase-space and symmetry of a virtual spherical water phantom. The symmetry also allows the number of doselets necessary for dose reconstruction to be reduced by a factor of ~250. In this work, 51 doselets were used. The SDM method mitigates the most computationally intensive part of this type of dose reconstruction – reading, weighting and summing dose matrices. The accuracy of the system was tested against MC calculations as well as our previously reported phase-space modulation (PSM) method using a series of open field and IMRT cases. The mean chi- and gamma-test 3% / 3 mm success rates of the SDM method were 98.6% and 99.5%, respectively, when compared to full MC simulation. The total calculation time was 96 seconds per treatment field on a single processor core.

Pre-treatment radiotherapy dose verification using Monte Carlo doselet modulation in a spherical phantom

1. **Introduction**

Modern radiotherapy treatment is a complex computer controlled process that involves multiple dynamically moving components. Techniques such as intensity modulated radiation therapy (IMRT) and volumetric modulated arc therapy (VMAT) are increasingly being implemented into clinical practice. The computerization of radiation therapy through these new technologies has resulted in a complex treatment system that may obfuscate the process, and despite considerable safety measures normally implemented, errors can slip into the treatment and undesirable dose can be delivered accidentally (Bogdanich 2010). Independent verification methods have therefore become essential to ensure that the treatment machines are delivering the expected radiation dose distributions to each patient. This has motivated mandates for individual dose verification as a part of quality assurance (QA) programs. There are two levels of such dose verification: pre-treatment and in-vivo dosimetry. In this paper we will focus on pre-treatment dose verification – a de-facto standard in clinical practice, though different methods vary extensively by their robustness, comprehensiveness and accuracy.

Common approaches to perform pre-treatment dose verification include the use of ionization chamber arrays (van Esch *et al* 2007, Saminathan *et al* 2010, Godart *et al* 2011) and diode arrays (Feygelman *et al* 2011). Such detector arrays provide effective 3D measurements, but tend to be expensive, heavy devices that make QA setup less than ideal. In addition, most of the current detector arrays are intended exclusively for use with coplanar treatment deliveries. Alternatively, direct analysis of machine log files can provide valuable information on treatment delivery (Litzenberg *et al* 2002 and Stell *et al* 2004), but does not reveal the impact of potential errors on the dose to a patient. Dynalog files in combination with Monte Carlo (MC) calculations, using the DOSXYZnrc code, have also been used for pre-treatment QA (Teke *et al* 2010) and provide accurate calculation of the dose delivered at the QA session. However, MC based systems tend to require significant expertise and long calculation times, limiting their use to relatively few clinics worldwide. In contrast, due to the near ubiquitous adaptation of electron portal imaging devices (EPIDs) by linac vendors, EPI-based verification techniques have become an active area of research (Ansbacher, 2006, Van Etmpt *et al* 2008, van Zijveld *et al* 2009, Yeo *et al* 2009, Mans *et al* 2010, Qian *et al* 2010). These techniques utilise linearity of EPI dose response (Popescu and Greer 2003) and have the advantage of using a device that is already present on the linac.

EPID dosimetry can be carried out using either transmission (with a patient or phantom in place) or non-transmission measurements. Both schemes can be used for either 2D fluence comparisons or 3D reconstructions to achieve verification. However, it was shown by Kruse (2010) that 2D verification of IMRT plans is insensitive and not sufficient to detect important dosimetric inaccuracies. Dose reconstructions in 3D tend to be more complex, since it is necessary to account for scattering that occurs in both the reconstruction volume and the imager itself. Solutions commonly assume coplanar treatment geometry (Ansbacher *et al* 2006) and therefore 65  only allow verification of coplanar treatments. While non-coplanar treatments are typically uncommon in IMRT and VMAT, they have demonstrated considerably improved dose conformity (Pugachev *et al* 2001, Wang *et al* 2005 and Llacer *et al* 2009) that is likely to be exploited increasingly in the near future. The dosimetric advantages are clear (normal tissue dose is spread over a larger volume), but implementation has generally been limited by the difficulties involved with treatment couch motion, which requires increased staff involvement, treatment time and more complex QA procedures. A modern generation of linacs with integrated robotic couch
70  motion reduces the difficulties associated with treatment delivery, but the problem of robust QA procedure remains.

In a previous work, our group presented an effective solution called the phase-space modulation (PSM) method (Berman et al 2010). In this technique, the (non-transmission) EPI signal was deconvolved to remove imager scatter and produce a fluence map. Each particle in a patient-independent phase-space was then weighted
75  ("modulated") according to the value of the fluence element intercepted by its projected path. The modulated phase-space was used for MC dose calculation to reconstruct the dose distribution that would have been delivered to a phantom or patient. The success of this strategy has led us to develop a new method that avoids using MC dose calculation during pre-treatment QA.

This paper presents a robust EPI-based pre-treatment verification technique capable of quickly
80  reconstructing 3D dose distributions from both coplanar and non-coplanar treatments. This will be referred to as spherical doselet modulation (SDM) method. The dose reconstruction is based on EPI measurements, and performed by modulating pre-calculated Monte Carlo doselets in a virtual spherical water phantom. The novelty of this technique is in essentially eliminating the statistical uncertainty of MC dose calculations by combining azimuthal symmetry in a patient-independent phase-space with the spherical symmetry of a water phantom to
85  derive patient-independent radial beamlet dose distributions (doselets). Only a small number of doselets are necessary for accurate dose reconstruction, compared to thousands when this symmetry is not exploited. To the best of our knowledge doselets of this type have not been used before, and they offer considerable advantages for precision calculations.

2. **Methods and materials**

90  *2.1    Phase-space sorting using azimuthal particle redistribution*
The Monte Carlo particle transport code BEAMnrc (Kawrakow and Walters 2006) was used for generating a planar phase-space upstream of all patient-dependent beam-shaping apertures in a 6MV Varian Clinac 21EX (Varian Medical Systems, Palo Alto, CA, USA) treatment head. An azimuthally symmetric circular source of electrons incident on the target was used. When transported through the azimuthally symmetric physical
95  geometry of the accelerator, it can be assumed that the resulting particle distribution is also symmetric. The phase-space was spatially discretized into beamlets, which, after having been transported through a phantom produced doselets in units of Gy per incident electron. In the context of this work, a beamlet refers to a single unique spatial region of the phase-space (no two beamlets may contain the same particles). Usually beamlets are defined to fill a Cartesian grid for a planar phase-space (Bush *et al* 2008a), but due to the symmetries utilised in this paper we use beamlets produced on a cylindrical grid defined by annular sectors. Due to rotational symmetry, all beamlets within an annulus are indeed dosimetrically equivalent. Therefore, using a modified azimuthal particle redistribution (APR) technique (Bush *et al* 2007), phase-space particles were redistributed into a single annular sector for each annulus, as illustrated in figure 1, dramatically increasing the particle density in the beamlet. For 1° azimuthal size of a sector, the particle density will increase by 360 times resulting in nearly 19-fold dose calculation uncertainty reduction (Spezi *et al* 2002). At the centre of the grid a circular region was defined as an additional beamlet, instead of being divided azimuthally.

For this work, the phase-space was divided into 50 annuli and one central circular sector within an outer circular boundary of radius 10.1 cm (projected to the isocentre). The central circular beamlet was 0.2 cm in diameter at the isocentre, the same as the radial thickness of each annulus. The particles in each annulus were compressed into a single corresponding annular sector beamlet of approximately 1.42° in angular size. This size corresponds to dividing each annulus into $N_\varphi$ = 254 sectors. The number of annuli and annular sectors were chosen by trial-and-error to achieve a reasonable balance of speed and accuracy for our system.

*2.2    Generation of patient-independent Monte Carlo doselets*

The phantom was generated as 20.25 x 20.25 x 20.25 cm$^3$ cube of air that contained a sphere of 10 cm in radius filled with water. Voxel dimensions of 0.125 x 0.125 x 0.125 cm$^3$ were used in the phantom.

For each of the beamlets defined by the phase-space sorting algorithm described above, DOSXYZnrc (Kawrakow and Walters, 2006) was used to generate doselets in this phantom. The doselets were initially calculated in Cartesian coordinates, and then converted into spherical coordinates (with origin at the centre of the water sphere) using tri-linear interpolation. In subsequent plan verification calculations, the plan isocentre was set to be at the centre of the spherical phantom at the source to axis distance (SAD). This allows us to define a spherical voxel system that, in the azimuthal plane, aligns with projected beamlets discussed in the previous section. That is, the radial divisions correspond 1-to-1 with the annuli, as do the azimuthal/annular sectors. This coordinate system is a key component in our method, as it removes the necessity for time-consuming dose interpolations during dose reconstruction (section 2.5) that could introduce interpolation artefacts.

To correspond with the beamlets, the spherical voxel system was defined with 0.2 cm radial divisions, $N_\varphi$ = 254 (~1.42°) azimuthal elements and $N_\theta$ = 134 (~1.34°) polar elements. Since the central circular beamlet was chosen as 0.1 cm in radius, the first radial division in the coordinate system was similarly reduced to 0.1 cm.

The entire process thus far is patient-independent and needs to be performed only once per phase-space (i.e. linac beam). One set of doselets can be re-used for all dose reconstructions so long as they require the same source model and spherical phantom. Thus high particle densities and small voxel resolution can be used during Monte Carlo dose calculation to attain very small statistical uncertainty, at no sacrifice to the speed of dose reconstruction.

*2.3    Construction of fluence maps from EPIs*

Similar to other pre-treatment verification methods using EPIs, patient-dependent portal images were collected prior to treatment for the spherical doselet modulation method. We followed the image acquisition strategy described by Ansbacher (2006). The EPID was positioned at a source-to-image distance (SID) of about 110 cm and then repositioned longitudinally in order to optimally include the area of all treatment fields. With a delivery rate of 400 MU/min, three images were acquired from dark, flood and 10 x 10 cm$^2$ calibration fields. The treatment field images were collected using the gantry angles specified in the treatment plan, and then corrected by subtracting the dark field and dividing by the flood field images. These images were normalised by the calibration field. Fluence maps were constructed by deconvolving the corrected EPIs using a kernel that accounts for scatter in the imager. This process was described in more detail in our paper on the phase-space modulation method (Berman *et al* 2010). For the purpose of the current work, the fluence map represents a weighting matrix with elements geometrically defined to be identical to the cylindrical grid used to generate the beamlets as projected to the isocentre.

*2.4    Doselet dose conversion to absolute units*

The phase-space source used for doselet generation was calibrated using a standard Monte Carlo simulation under calibration conditions (Popescu *et al* 2005). The normalisation factor $D_{10\times10}$ was derived as the Monte Carlo dose (in Gy/e$^-$) at the central axis of a 10 x 10 cm$^2$ field at 10 cm depth in water. The tissue maximum ratio (*TMR*) value (in Gy) was used as a reference dose at these calibration conditions. The SDM method requires two further factors for absolute dose calibration: the number of azimuthal sectors, $N_\varphi$, to account for increased particle density from APR, and the monitor units used for producing the 10 x 10 cm$^2$ calibration EPI, $MU_{cal}$. The fluence map elements $w_{\varphi,r}$ that were derived from EPIs already account for the monitor units as well as monitor chamber backscatter (Zhu *et al* 2009) from each field. The conversion of the doselet voxel dose $D_{\varphi,r}$ from the relative units of Gy/e$^-$ to absolute dose $D_{\varphi,r}^{abs}$ in Gy then becomes

$$D_{\varphi,r}^{abs} = D_{\varphi,r} \cdot w_{\varphi,r} \frac{MU_{cal} \cdot TMR}{D_{10\times10} \cdot N_\varphi}.$$

*2.5    Re-constructing 3D dose in a sphere*

Once we have patient-dependent fluence maps derived from EPIs, dose reconstruction can begin, as outlined by the flowchart in figure 2. Since the doselets are generated only for one annular sector per annulus from the phase-space, each doselet must be azimuthally rotated and re-used $N_\varphi$ times to fill an annulus. Recall that elements on the azimuthal plane of the spherical coordinate system of the dose distributions correspond with beamlet divisions. This means that the dose from a given annulus can be collected by: (1) repeatedly re-indexing the azimuthal coordinates of the corresponding doselet, (2) multiplying the original dose values by the fluence element weighting factor and (3) adding the dose in each re-indexed voxel to the cumulative dose matrix for the field (figure 2). There is one exception to this process - the central doselet was defined corresponding to only a single element in the fluence map, so no rotation is necessary. The doselet rotation/modulation/summation procedure is the most computationally intensive part of the SDM method, since the entire 3D dose distribution for each doselet must be re-indexed, scaled and summed $N_\varphi$ times. Once every fluence map element has been used, the sphere will be filled with a patient-specific dose distribution from one treatment field. Exactly the same procedure is done to derive the dose distributions for each treatment field in the plan.

Gantry angle rotation is applied to the dose distribution for the field using the values acquired from the portal image DICOM file. Collimator rotations are accounted for implicitly in the EPI as the image captures delivered particle fluence. Couch rotation, however, is not executed during verification measurement to avoid collisions. Instead, the couch angle is read from the plan DICOM file and performed at the dose reconstruction time. Since the sphere is centred at the isocentre, gantry and couch rotations could be simply implemented by further re-indexing the dose coordinates along the polar and azimuthal directions, respectively. This technique would provide fast dose rotations, but is a "nearest neighbour" approximation. Since this rotation is necessary only once per field, the contribution to the overall calculation time is small. For this reason, we chose to use more advanced interpolations: the dose for each field is first converted back into Cartesian coordinates, and then the gantry and couch rotations (as delivered) are induced using cubic spline interpolation. The Cartesian coordinate system is necessary for comparison of the reconstructed dose with the planned dose distribution. The conversion from spherical to Cartesian coordinates is performed using tri-linear interpolation. After the above procedure has been completed for each field, the contributions are simply summed to obtain the total dose.

*2.6    Testing the spherical doselet modulation method*

The spherical doselet modulation method has been tested against standard Monte Carlo simulations and the PSM method. Since the purpose of these tests was to illustrate the accuracy capability of the SDM method, all simulations shared identical treatment parameters. In other words, the same gantry angles, MUs, etc. were used in all cases (rather than comparing planned versus delivered). The PSM and MC benchmark utilized

DOSXYZnrc for dose calculation in a cubic phantom with 2.5 x 2.5 x 2.5 mm$^3$ voxel resolution containing a 10 cm radius water sphere. After dose reconstruction in spherical coordinates, the SDM results were also interpolated onto this phantom. All methods used the same phase-space source scored just above the secondary collimators from a BEAMnrc model of a 6MV Varian Clinac 21EX (Varian Medical Systems, Palo Alto, CA, USA). This model has been validated previously (Bush *et al* 2007, Gagne and Zavgorodni 2007, Bush *et al* 2009, Bush *et al* 2011). In the benchmark MC calculations, secondary collimator simulation was performed using BEAMnrc. In the other two methods, beam-shaping information is implicit in the portal images. The energy cut-offs (PCUT and ECUT) in DOSXYZnrc for photons and electrons were 0.010 MeV and 0.700 MeV, respectively. In every plan the isocentre was set to the centre of the spherical phantom at 100 cm SAD. Computations were performed on the VIMC CPU cluster that includes three compute nodes, each with four AMD Opteron 2.1 GHz 16-core processors, 192 GB DDR3 RAM and 7200 RPM SATA hard drives. The benchmark and PSM simulations utilized the Vancouver Island Monte Carlo (VIMC) framework for streamlined dose calculation (Zavgorodni *et al* 2007, Bush *et al* 2008b). The SDM method has now also been integrated into this framework.

Open fields of size 1 x 1, 3 x 3, 5 x 5, 10 x 10, 15 x 15 and 20 x 20 cm$^2$ were considered, along with 6 clinical head and neck IMRT cases (four brain, one larnyx and one left tonsil). One of the brain cases included a non-coplanar field.

3. Results

The total calculation time for dose reconstruction per EPI was 96 seconds using a single 2.1 GHz processor. This time includes applying portal image calibration corrections, deriving a fluence map, rotating/modulating/summing doselets and applying gantry/couch rotations for a single EPI. The bulk of this time (81 seconds) is spent rotating, modulating and summing doselets.

Depth dose and cross-beam profiles for a variety of open field sizes are provided in figures 3 and 4, respectively. The percentage difference between the SDM method and MC benchmark was plotted in these figures, relative to the maximum dose in the benchmark. In these relatively uniform cases, systematic spherical ringing artefacts of the order of 2% can be observed throughout the phantom – these have been attributed to the tri-linear interpolations used for conversion between Cartesian and spherical coordinate systems. The SDM method also exhibits systematic errors in penumbra regions due to beamlet/doselet discretization. This motivates the choice of chi- and gamma-index test (Low *et al* 1998 and Bakai *et al* 2003) criteria as 3% / 3 mm above 80% and below 20% of the maximum dose, limited to 8 cm radius from the sphere centre.

The six clinical IMRT head and neck cases tested achieved good agreement with MC. The results of these tests are provided in table 1, and isodose images for the "Brain4" case (with two non-coplanar fields) are shown in figure 5. The mean chi- and gamma-test success rates were 98.6% and 99.5%, respectively, when compared with the standard MC simulation benchmark. Absolute root mean square (RMS) dose differences in the regions of interest were no larger than 2.3% in all cases.

## 4. Discussion

Existing dose verification methods that use convolution to account for EPI and phantom scatter commonly use an invariable scatter kernel, resulting in calculation accuracy loss off-axis and at-depth (Ansbacher, 2006). The SDM method avoids these problems by using Monte Carlo calculated doselets that inherently account for fluence spectral changes as photons get transported through the linac head and phantom. Since the doselets are patient-independent and only calculated once, it is also possible to achieve very low statistical uncertainty using long initial simulation times. Further use of the doselets for QA calculations does not require any Monte Carlo simulations to be performed, so the necessary computational overhead and expertise is minimized. This makes it possible to share the doselet database between institutions, where a Monte Carlo program may be difficult to establish. The SDM method also has the advantage of minimal experimental set-up, requiring only measurements in air using an EPID. Non-coplanar treatments are supported even in cases where the couch would collide with the imager, since the in-air measurements do not require rotation of the couch (rather, it is sufficient to simply rotate the reconstructed field dose distribution).

For this work, algorithm parameters were selected to provide a reasonable trade-off between dose reconstruction speed and accuracy. For different hardware configurations and accuracy requirements, these parameters could vary. The most important of these is the dose voxel resolution in the spherical coordinate system. The dose reconstruction time is approximately inversely proportional to the voxel volume and larger voxels could have been chosen. To mitigate accuracy dependency on voxel dimensions alternative interpolation methods such as cubic spline could be considered. Additionally, modeling particle transport directly in spherical coordinates would eliminate the need for Cartesian to spherical interpolation (though conversion from spherical to Cartesian coordinates after dose reconstruction would still be necessary).

The spherical geometry of the verification phantom used in this study is different to the commonly used cylindrical water phantoms. The reasons for the more conventional cylindrical shape of verification phantom include similarity to human geometry, rotational symmetry that simplifies dose calculations for coplanar treatments, and ease of physical construction when used in experimental verifications. In our technique the ease of construction is irrelevant, as the phantom is "virtual" and does not actually need to be manufactured. Also, so long as the attenuation of the radiation field is not excessive (nor insufficient) near the regions of interest, the shape of the phantom does not need to mimic the human form to achieve meaningful dose verification. A spherical surface simplifies the dose reconstruction algorithm by avoiding the necessity of corrections for irregularities (unlike the edges of a cylinder in non-coplanar treatments). For the SDM method, the spherical shape was essential for the high efficiency of the technique. In the case of large field sizes or treatment areas, it might be necessary to increase the radius of the sphere to encompass the entire field, and this should probably be combined with corresponding reduction of the phantom density to avoid unrealistic beam attenuation and maintain the sensitivity of the method to capturing treatment errors. We leave the consideration of this idea for future work.

## 5. Conclusions

The SDM method has been shown to provide effective 3D dose verification using EPIs in conjunction with pre-calculated MC doselets. High efficiency calculations were achieved by azimuthally compressing a patient-independent phase-space into beamlets with extremely high particle density. Only a small number of doselets in a spherical phantom were required to perform accurate dose reconstruction. This novel strategy exploited rotational symmetries to mitigate the most computationally intensive part of dose reconstruction – reading, weighting and summing dose matrices. In the open field and IMRT treatment cases used to test the accuracy of the method, the mean chi-test success rate of the SDM method was 98.6% with a calculation time of 96 seconds per EPI on a single processor core.

Pre-treatment radiotherapy dose verification using Monte Carlo doselet modulation in a spherical phantom

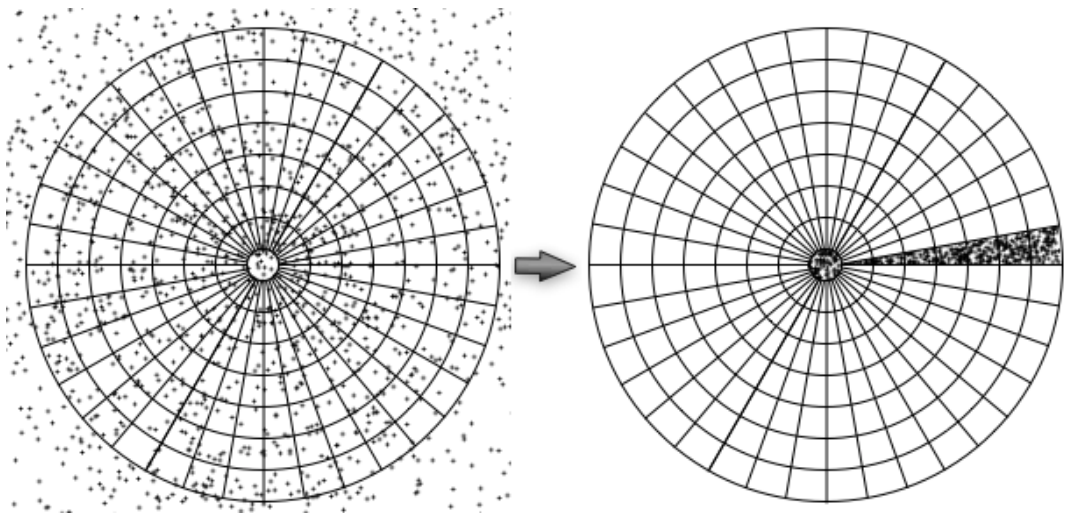

335  Figure 1. This figure visually illustrates the beamlet generation procedure. Particles in the phase-space were sorted using a cylindrical grid. Azimuthal particle redistribution (APR) was used to rotate particles into a single annular sector per annulus. In the central circular beamlet, recycling with APR is performed to similarly increase the particle density.

Pre-treatment radiotherapy dose verification using Monte Carlo doselet modulation in a spherical phantom

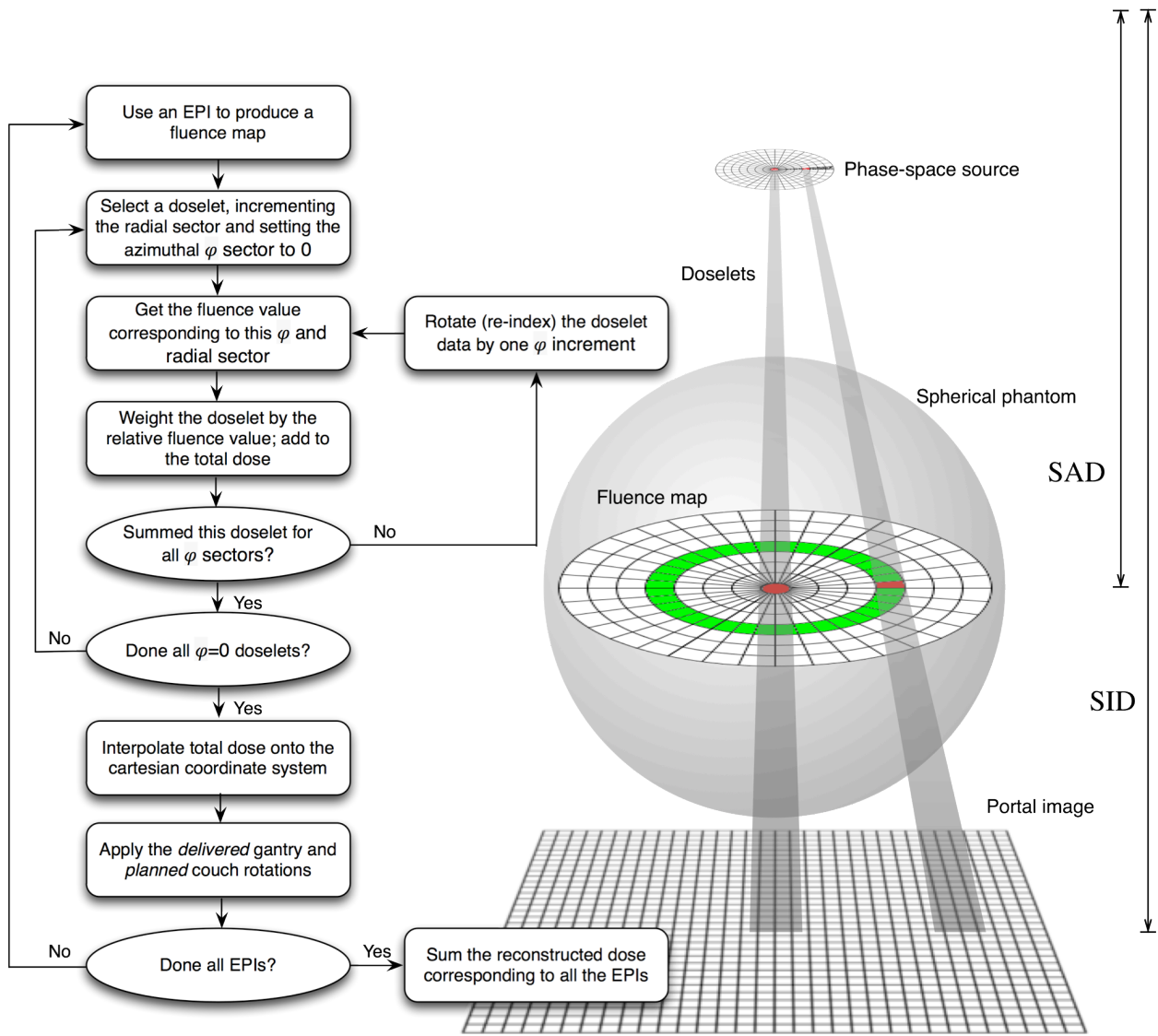

Figure 2. A flowchart and diagram illustrating how beamlets from the phase-space source were used to calculate doselets in a spherical phantom (not to scale). In this process, an electronic portal image (EPI) was deconvolved to produce a fluence map projected to the source-to-axis distance (SAD). The fluence map elements corresponding to two doselets are highlighted in the figure.

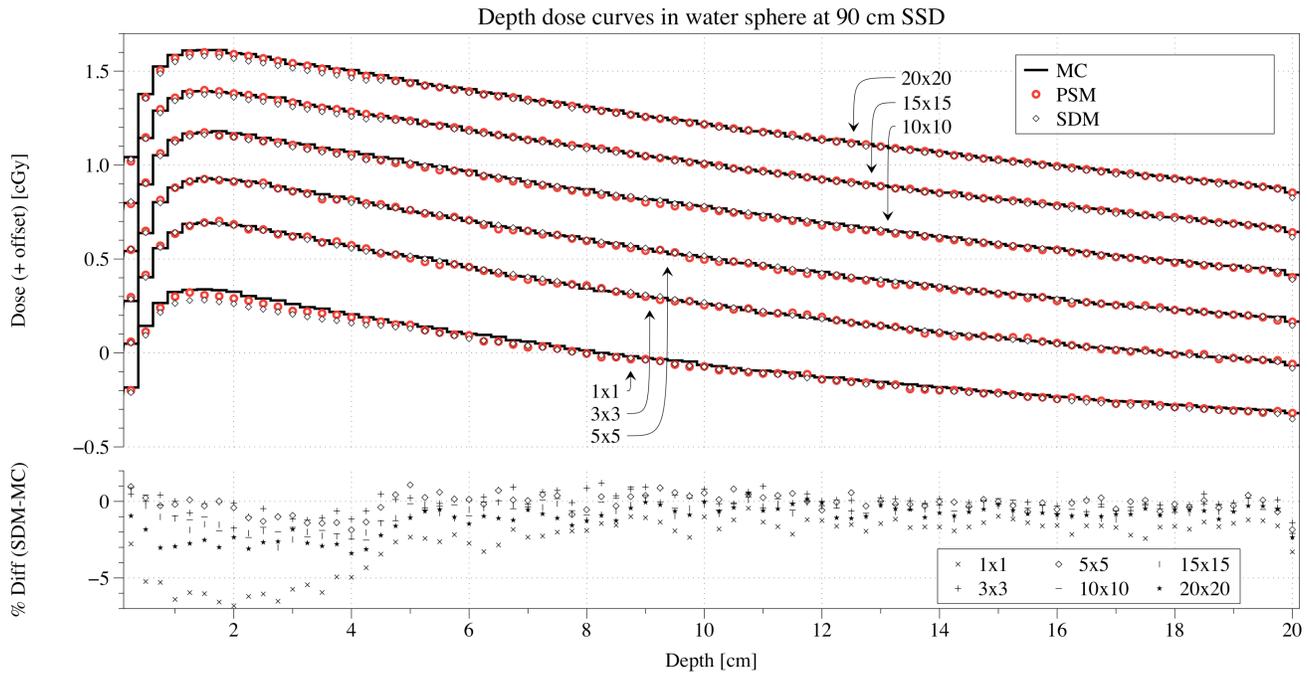

Figure 3. Depth dose curves are shown for a variety of field sizes for the spherical doselet modulation (SDM) method, phase-space modulation (PSM) method, and a standard Monte Carlo (MC) simulation. Curves were artificially offset by 0.2 cGy increments for clarity, except for the 10 x 10 cm$^2$ curve. The percentage differences between the SDM method and MC simulation are shown, relative to the maximum MC dose. All simulations were performed in a spherical 10 cm radius water phantom at 90 cm SSD.

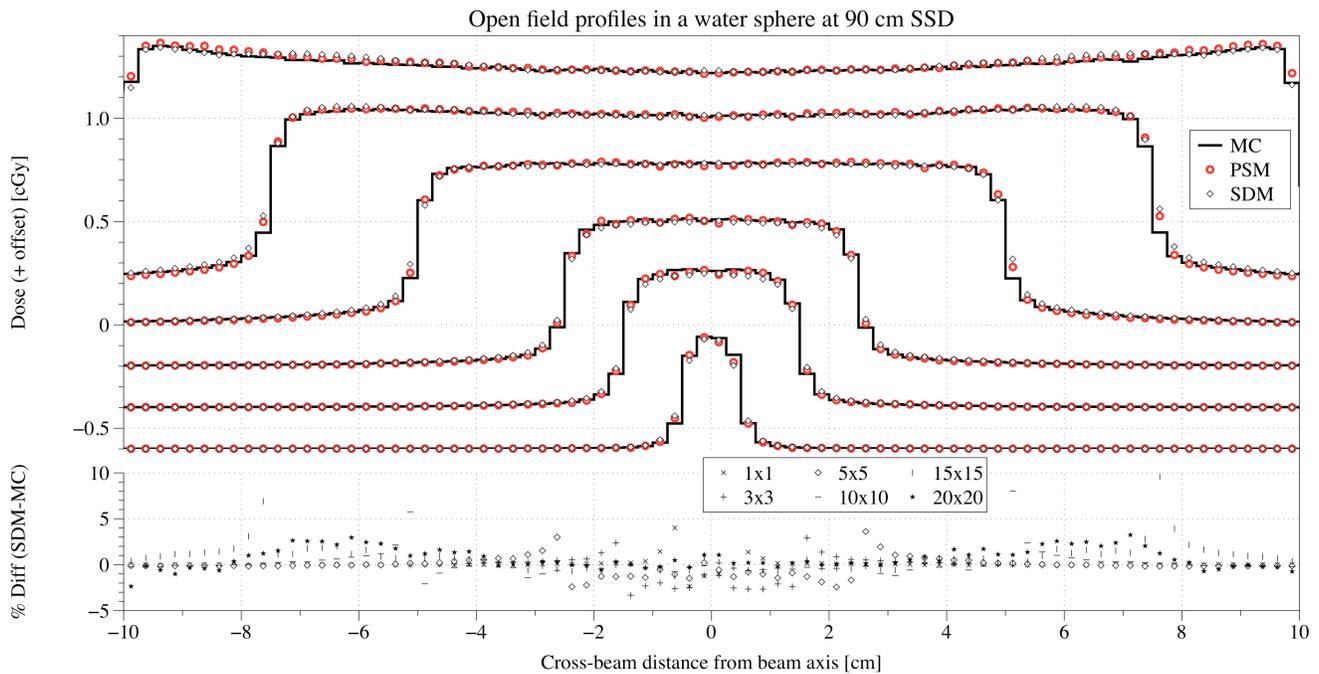

Figure 4. Cross-beam profile curves at 10 cm depth are shown for a variety of field sizes for the spherical doselet modulation (SDM) method, phase-space modulation (PSM) method, and a standard Monte Carlo (MC) simulation. Curves were artificially offset by 0.2 cGy

350  increments for clarity, except for the 10 x 10 cm$^2$ curve. The percentage differences between the SDM method and MC simulation are shown, relative to the maximum MC dose. All simulations were performed in a spherical 10 cm radius water phantom at 90 cm SSD.

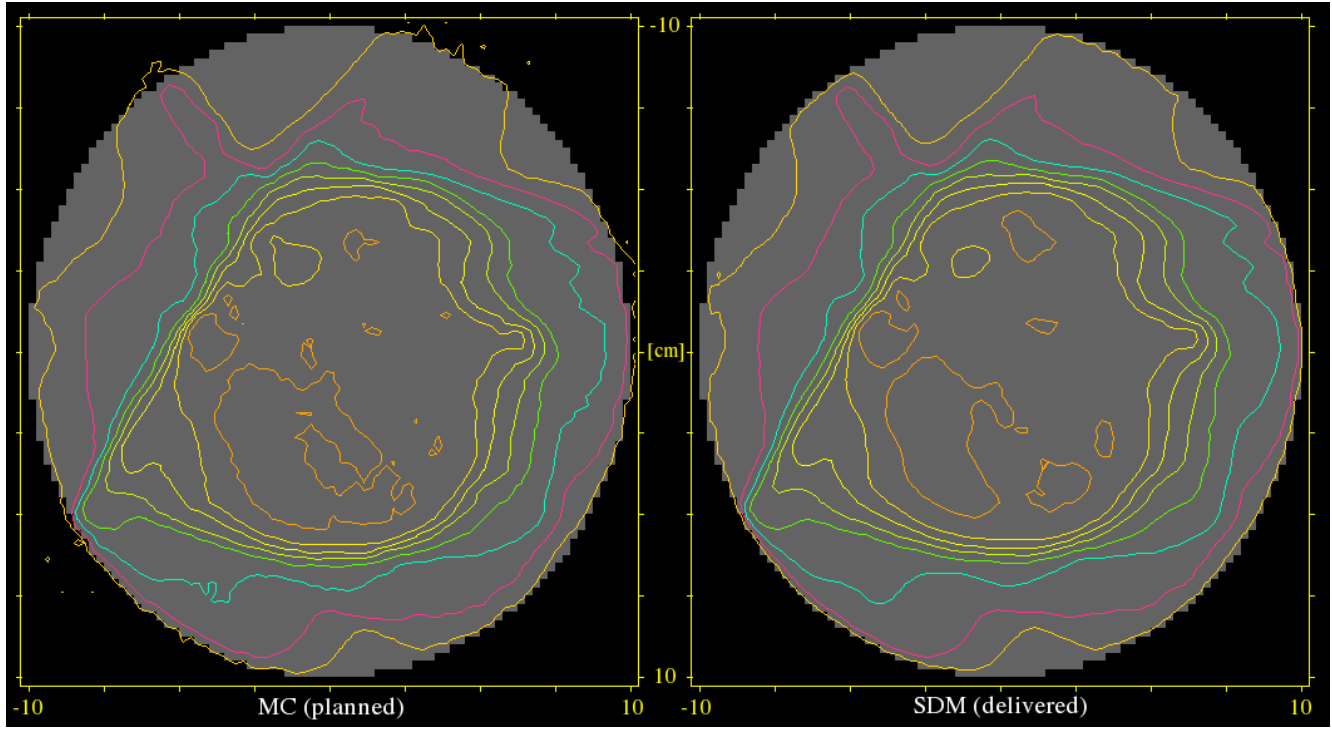

Figure 5. Isodose lines for a non-coplanar IMRT treatment (in the coronal plane through the centre of the sphere) showing standard Monte Carlo calculation (left) and spherical doselet modulation (SDM) method (right). Isodose lines start at 10% of the maximum dose and
355  increment in 10% intervals up to 90%.

| Plan | Gamma Success (%) | | Chi Success (%) | | RMS Diff (%) | |
|---|---|---|---|---|---|---|
| | SDM | PSM | SDM | PSM | SDM | PSM |
| 1x1 | 100. | 100. | 99.9 | 99.9 | 0.4 | 0.3 |
| 3x3 | 100. | 100. | 99.5 | 99.7 | 0.7 | 0.5 |
| 5x5 | 99.9 | 100. | 99.4 | 99.8 | 1.0 | 0.6 |
| 10x10 | 100. | 99.9 | 96.5 | 99.3 | 2.3 | 1.3 |
| 15x15 | 99.5 | 99.8 | 97.1 | 99.4 | 1.9 | 1.4 |
| 20x20 | 97.9 | 100. | 96.0 | 100. | 1.7 | 0.9 |
| Larynx | 99.4 | 99.0 | 99.0 | 97.6 | 1.4 | 1.7 |
| LT Tonsil | 100. | 100. | 100. | 100. | 1.2 | 1.0 |
| Brain1 | 99.3 | 99.8 | 100. | 99.3 | 1.9 | 1.5 |
| Brain2 | 98.2 | 100. | 97.1 | 99.6 | 1.8 | 1.3 |
| Brain3 | 100. | 99.2 | 99.9 | 98 | 1.0 | 1.6 |
| Brain4 | 100. | 100. | 99.6 | 99.7 | 1.3 | 1.3 |

Table 1. Results comparing the spherical doselet modulation (SDM) method and phase-space modulation (PSM) method to a standard Monte Carlo calculation. All simulations were performed in a 10 cm radius spherical phantom. Chi- and gamma-index tests used 3% / 3 mm criteria contained to voxels within 8 cm radius of the sphere centre above the 80% and below the 20% isodoses. The root mean square (RMS) of dose differences is also shown.